\documentclass[aps,prd,twocolumn,showpacs,floatfix,superscriptaddress,preprintnumbers]{revtex4}

\usepackage{epsfig}

\usepackage[dvips]{color}
\definecolor{Black}{named}{Black}
\definecolor{Red}{named}{Red}
\definecolor{Blue}{named}{Blue}

\def\lsim{\raise0.3ex\hbox{$\;<$\kern-0.75em\raise-1.1ex\hbox{$\sim\;$}}}
\def\gsim{\raise0.3ex\hbox{$\;>$\kern-0.75em\raise-1.1ex\hbox{$\sim\;$}}}
\def\eps{\varepsilon}
\def\theta{\vartheta}

\def\dcp{\delta_{\rm CP}}

\newcommand{\be}{\begin{equation}}
\newcommand{\ee}{\end{equation}}
\newcommand{\bea}{\begin{eqnarray}}
\newcommand{\eea}{\end{eqnarray}}

\begin{document}

\preprint{IFIC/06-22}

\title{High energy neutrino yields from astrophysical sources I:
Weakly magnetized sources}

\author{M.~Kachelrie\ss}
\affiliation{Institutt for fysikk, NTNU Trondheim, N--7491 Trondheim,
  Norway}

\author{R.~Tom\`as}
\affiliation{AHEP Group, Institut de F\'{\i}sica Corpuscular -
  C.S.I.C/Universitat de Val\`encia\\
Edifici Instituts d'Investigaci\'o, Apt. 22085, E-46071 Val\`encia, Spain}


\begin{abstract}
We calculate the yield of high energy neutrinos produced in astrophysical 
sources with negligible magnetic fields varying their interaction
depth from nearly transparent to opaque. We  
take into account the scattering of secondaries on background photons as 
well as the direct production of neutrinos in decays of charm mesons. 
If multiple scattering of nucleons becomes important, the neutrino spectra 
from meson and muon decays are strongly modified with respect to transparent 
sources. Characteristic for neutrino sources containing photons as
scattering targets is a strong energy-dependence of the ratio $R^0$ of 
$\nu_\mu$ and $\nu_e$ fluxes at the sources, ranging from
$R^0=\phi_\mu/\phi _e\sim  0$ below threshold to $R^0\sim 4$
close to the energy where the decay length of charged 
pions and kaons equals their interaction length on target
photons. Above this energy, the neutrino flux is
strongly suppressed and depends mainly on charm production.
\end{abstract}


\maketitle

\section{Introduction}

Experimental high energy neutrino physics has become one of the most active
areas of astroparticle physics, offering among others the prospect of
identifying the sources of ultra-high energy cosmic rays~\cite{reviews}. High 
energy neutrinos from astrophysical sources are the decay products of
secondary mesons produced by scattered  high energy protons on background
protons or photons.  The classic example are the so-called cosmogenic
or GZK neutrinos produced in scatterings of extragalactic ultra-high
energy cosmic rays on cosmic microwave photons during
propagation~\cite{GZKnu}. Two different kind of bounds on
high energy neutrino fluxes exist: The cascade or EGRET limit uses bounds on
the diffuse MeV-GeV photon background to limit the energy transferred to
electromagnetically interacting particles that are produced unavoidably
together with neutrinos~\cite{casc}. The cosmic ray upper
bounds of, e.g., Refs.~\cite{WB,MPR} use the observed ultra-high
energy cosmic ray flux to limit
possible neutrino fluxes. The latter limit assumes that all neutrino sources
are transparent to hadronic interactions and thus at least neutrons
can escape from the source region without interactions. Dropping the
assumption of transparent sources and hidding the acceleration region by 
sufficient material absorbing ultra-high energy cosmic rays 
allows one to avoid the cosmic ray 
limits~\cite{TZ77,BG81,St,Berezinsky:2000bq}. One might therefore
speculate that large neutrino fluxes at high  
and ultra-high energies might be produced in opaque sources, as they 
are needed, e.g., to perform neutrino absorption spectroscopy~\cite{spec}.

In this work, we calculate the flux of high energy neutrinos produced as 
secondaries in astrophysical sources. We consider sources with
interaction depth  
ranging from nearly transparent to opaque. In contrast to most earlier
investigations~\cite{Y,earlyI,earlyII}, 
we put emphasis on sources with such high 
densities that magnetic fields can be neglected and 
multiple scatterings are important. As a result, the neutrino spectra 
from meson and muon decays are strongly modified with respect to transparent 
sources. Since in most astrophysical environments the depth for $p\gamma$ 
is above threshold much larger than for $pp$ interactions, 
we consider photons in the source as target material.
In Sec. 2, we discuss the relevant production processes and
present the neutrino yields from a single source. We discuss also
briefly the consequences of the derived energy dependence of the
neutrino yields on the expected high energy neutrino fluxes from
astrophysical sources. In Sec.~3, we examine the neutrino
flavor composition at the source as well as the effect of neutrinos
oscillations on the flavor composition at the detector and find
significant deviations from the canonical flavor ratio expected from
pion decay. Finally, we summarize our results in  Sec.~4.

\section{Neutrino production processes and yields}

\subsection{Simulation and particle interactions}

We idealize a hidden neutrino source as an homogeneous slab filled
with photons in which high energy protons are  injected. The probability
${\cal N}$ that a particle travels in the slab the distance $\Delta L$ 
without scattering or decay is given by 
\be
 {\cal N} = \exp\left(-\int_{L}^{L+\Delta L} 
          \frac{{\rm d}l}{(l_{1/2} + l_{\rm int})} \right)  \,,
\label{delta}
\ee
where $l_{1/2}$ and $l_{\rm int}$ are its decay and interaction length,
respectively. Their relative value determines the fate of
the particle, either it decays or scatters. 

In our Monte Carlo simulation, we track explicitly all secondaries
($N,\pi^\pm,~K^\pm,~K^0_{L,S}$) for which the interaction rate is 
non-negligible compared to their decay rate. We determine the
multiplicity and energy spectra of light secondary particles produced
in scattering processes with a modified version of SOPHIA~\cite{sophia}. 
For the case of hadrons containing charm quarks, we employ
HERWIG~\cite{HERWIG} to determine the differential energy
spectra of prompt charm neutrinos, while we estimate the total cross
section for charm production from Ref.~\cite{charmbranching}.
To fix the pion-photon and kaon-photon interactions, we use the
following simple recipe~\cite{mm}: We determine first
the maximum of the resonant production cross section  in the
Breit-Wigner approximation of $\rho$ and $K^\ast$,
respectively, and compare it to
$p+\gamma\to\Delta\to$ all. Then we rescale the proton-photon cross
section by the ratios of the maxima to obtain the pion-photon and
kaon-photon cross sections in the non-resonant region.
Finally, the interaction length $l_{\rm int}$ is calculated from
\be
 l_{\rm int}^{-1} = \frac{1}{2\Gamma^2}
 \int_{\eps_{\rm th}}^{\infty} d\eps' \:\sigma_{h\gamma}(\eps')
 \int_{\eps_{\rm th}/2\Gamma}^{\infty} d\eps \:\frac{n(\eps)}{\eps^2} \,, 
\ee
where $\Gamma$ is the Lorentz factor of the hadron in the lab frame,
$\eps_{\rm th}$ is the threshold energy of the relevant reaction in
the rest system of the hadron, and $n(\eps)$ the energy distribution of
photons with energy $\eps$ in the lab frame.

\subsection{Phenomenological characterization of the sources}

In the following, we do not consider any particular model of neutrino 
sources, but characterize instead the sources in a
phenomenological way. This allows us to perform a systematic
analysis of different sources according to their thickness, 
from nearly transparent ones where the accelerated protons hardly
scatter once with the surrounding material to completely opaque sources
where neither high energy cosmic rays nor high energy photons
can escape. In our idealization of a neutrino source as an homogeneous
slab filled with photons, each source is fully characterized by its
length $L$ and the photon distribution $n(\eps)$. Potential sources of photon
backgrounds are manifold, but we restrict ourselves in this work to
the most important example, a background of thermal photons at
temperature $T$. 

A more severe restriction of the current work is that
we consider only sources with negligible magnetic
fields. Thus we require that i) energy losses due to synchrotron losses
are much smaller than due to interactions and that ii) deflections of
charged particles in magnetic fields are small compared to the size of
the source. This limitation allows us to stress more clearly the
salient features of opaque sources. In a subsequent work, we shall
consider the generic case of sources with arbitrary interaction depth
and magnetic fields~\cite{II}.

For the discussion of our numerical results it is useful to introduce
following Ref.~\cite{Y} several dimensionless quantities. The first
one is the ``interaction depth'' of nucleons defined analogously to the
optical depth as the ratio $\tau=L/l_{\rm int}$ of the size 
$L$ of the source to the interaction length $l_{\rm int}$ of
nucleons. This ratio determines, if most of the protons leave the
source without interactions ($\tau\ll 1$ or ``transparent source''),
or if multiple-scattering of nucleons is important and mesons are
efficiently produced ($\tau\gg 1$ or ``opaque source''). For the
illustration of our numerical results, we determine in the following
$\tau$ via $l_{\rm int}=1/(n_\gamma\sigma)$ with $\sigma=0.2$~mb as
reference cross section.

As second parameter we introduce the dimensionless energy $x\equiv
E\omega/m_p^2$, where $\omega=1.6T$ denotes the energy of the maximum of the
Planck distribution and $m_p$ the proton mass. This parameter allows
one to express the different yields in the form of a universal function,
which in the case of transparent sources does not explicitly depend
on the temperature. Finally, it is useful to introduce the critical
dimensionless energy $x_{\rm cr}=E_{\rm cr}\omega/m_p^2$, 
where $E_{\rm cr}=m/(\sigma n\ell_0)$ is defined as the energy at
which the decay length  $\ell_{1/2}=\Gamma c\tau_{1/2}=\Gamma l_0$ of a
meson with life-time $\tau_{1/2}$ and mass $m$ is equal to its
interaction length $\ell_{\rm int}$.  Using charged pions as
reference particles, $x_{\rm cr}$ depends numerically as
$x_{\rm cr}\approx 2\times 10^{10}/(T/{\rm K})^2$ on the temperature of
thermal photons.
Thus, the parameter $\tau$ determines the fraction of nucleons that
scatter on photons, while the parameter $x$ controls if mesons
mainly decay ($x\ll x_{\rm cr}$) or scatter ($x\gg x_{\rm cr}$).

\begin{figure}[ht]
\begin{center}
\includegraphics*[width=0.45\textwidth,angle=0,clip]{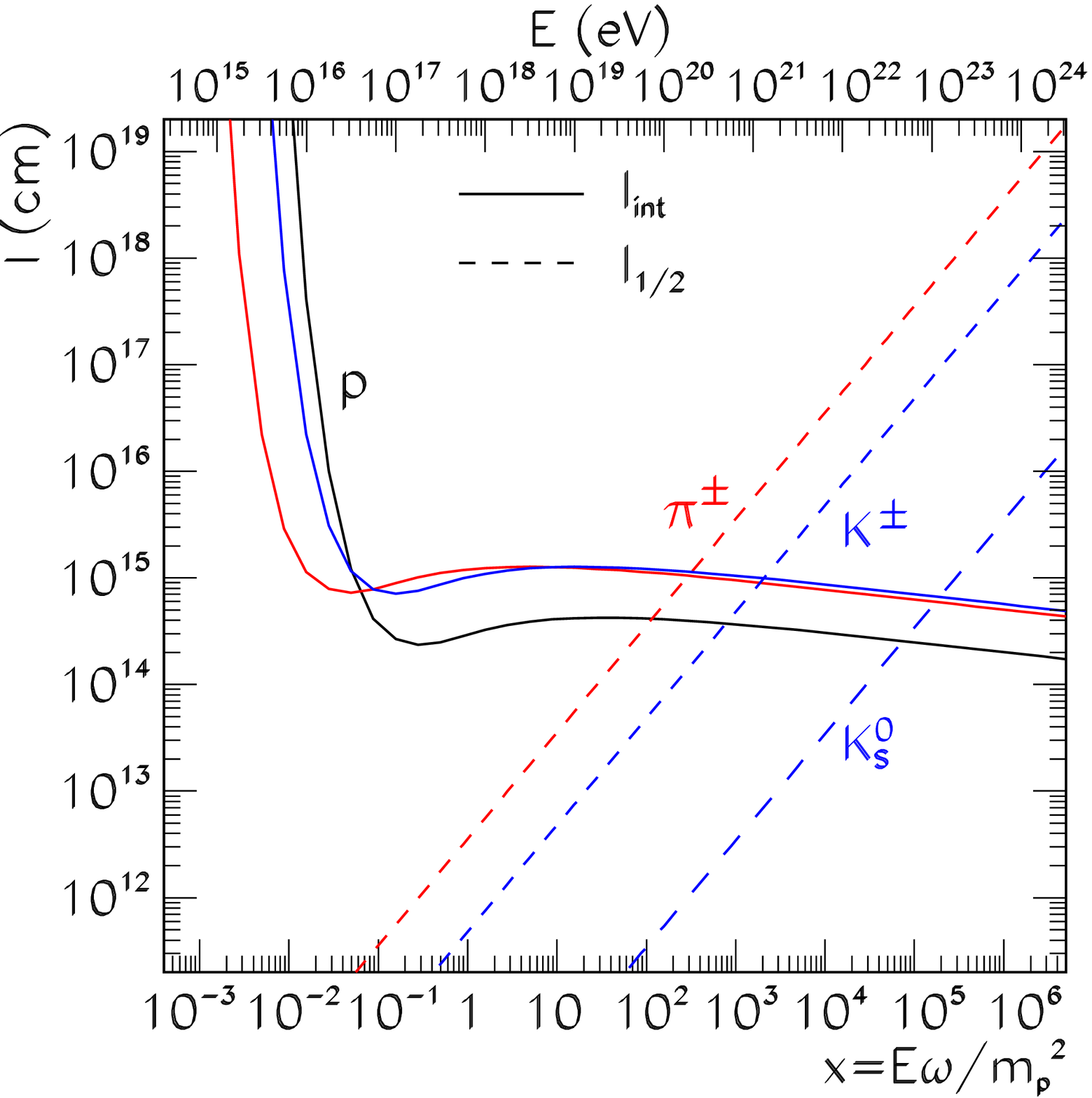}
\includegraphics*[width=0.45\textwidth,angle=0,clip]{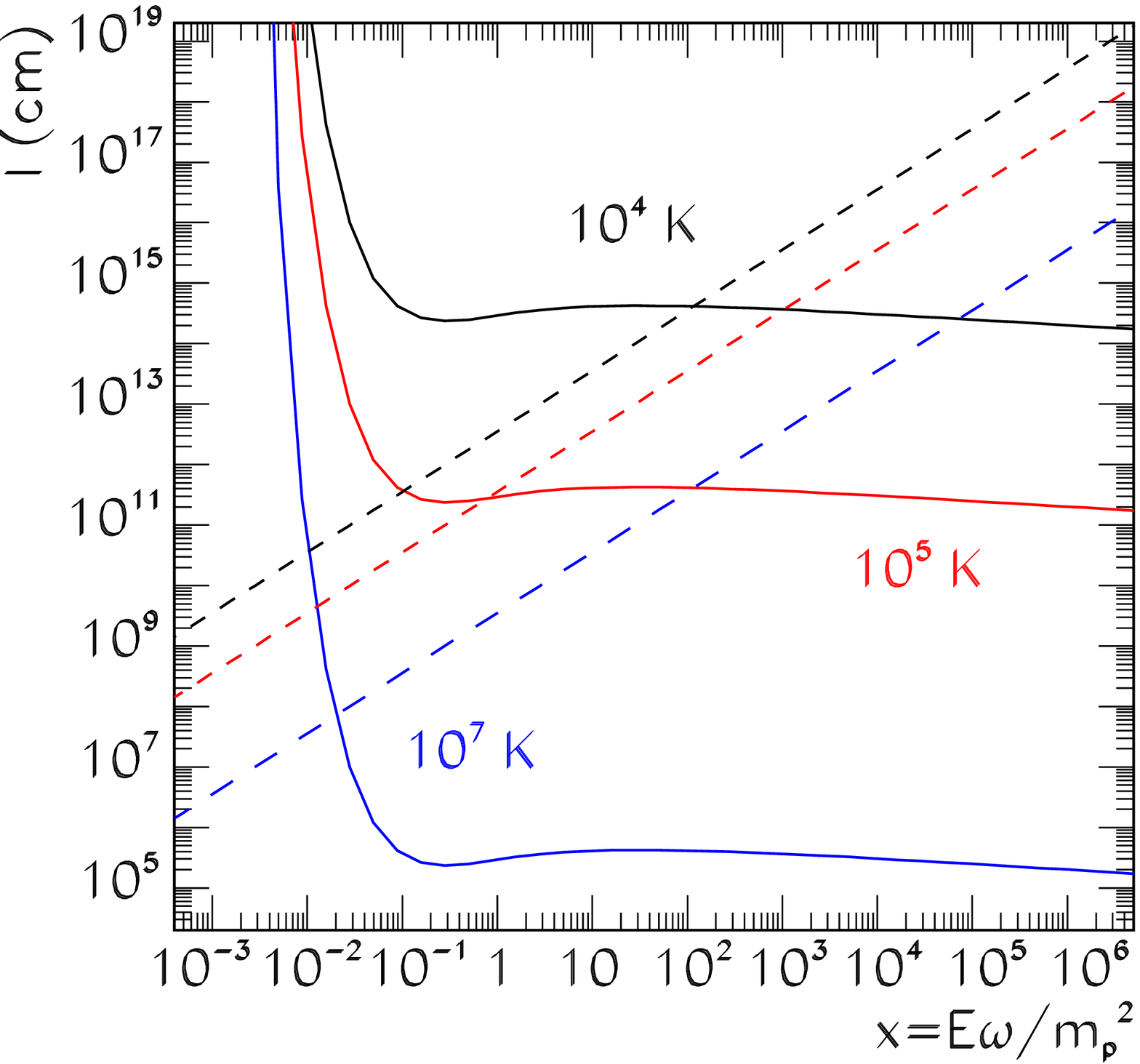}
\caption{\label {f_T4-5-7} 
(Color online) Top: Interaction length (solid lines) for proton in black, charged
pions in red (gray) and charged kaons in blue (dark gray) with 
thermal photons at a temperature $T=10^4$~K, together with the decay
length (dashed lines) for $\pi^\pm,~K^\pm$ and $K^0_S$.
Bottom: Interaction length (solid lines) for proton and decay length
(dashed lines) of charged pions
for sources at $T=10^4$~K in black ,  $T=10^5$~K in red (gray),
and  $T=10^7$~K in blue (dark gray).}
\end{center}
\end{figure}

Before we discuss the details of the different spectra, we briefly
comment on the main characteristics of  hadron-photon 
interactions. In the upper panel of Fig.~\ref{f_T4-5-7}, we show the
interaction length of protons, pions and kaons in a thermal bath of
photons with temperature $T=10^4$~K  together with their decay length
as function of the dimensionless energy $x$ (bottom axis) or
their energy (upper axis). The fate of a hadron is characterized by
two important energies: The threshold energy at $x_{\rm th}\sim 0.1$,
below which photo-meson reactions are exponentially suppressed, and
the critical energy $x_{\rm cr}$,  above which most mesons will scatter
before decaying. If the decay length and the
interaction length cross inside the source, $l_{1/2}=l_{\rm int}<L$,
the effect of multiple
scattering has to be taken into account. Main consequence is a 
strong suppression of the high energy neutrino flux produced in meson 
decays for $x\gsim x_{\rm cr}$. The shorter life-time 
of kaons with respect to pions as well as their larger mass 
 leads to a shift of their critical energy, i.e.\ 
$x^K_{\rm cr}>x^\pi_{\rm cr}$.

In the lower panel of Fig.~\ref{f_T4-5-7}, we show the interaction
and decay length of hadrons  as a function of $x$ for
sources with different temperature $T$. 
As expected from the temperature dependence of $l_{\rm int}$ and the
definition of $x$, the critical energy $x_{\rm cr}$ does not only
depend on the meson considered but also on the temperature: 
$x_{\rm cr}$ scales simply as $x_{\rm cr}\propto 1/T^2$ as long as
$x_{\rm cr}$ is above threshold, $x_{\rm cr}\gsim x_{\rm th}\sim 0.1$.  
Thus, multiple scattering starts to become important at lower energies
as the source temperature increases. On the other hand, the dimensionless 
threshold energy $x_{\rm th}$ is nearly independent of $T$. Therefore
in the case of opaque sources
the range with unsuppressed neutrino fluxes, $0.1 \sim x_{\rm th}\lsim
x \lsim x_{\rm cr}\propto 1/T^2$, shrinks for increasing $T$
and becomes a narrow peak around $x_{\rm th}$ for $T\gsim 5\times
10^5$~K. 
On the other hand, large neutrino fluxes at energies
high enough to perform neutrino spectroscopy, $E_{\rm
  cr}\gtrsim10^{21}$, can be obtained only in 
 sources at low temperatures, $T\lsim 5\times
10^3$~K. If, additionally, one requires the source to be opaque 
so that the high-energy cosmic ray flux is suppressed, 
e.g. $\tau>10$,
then a second
condition for a suitable source is a large extension, $L\gtrsim 3\times
10^{27}/(T/{\rm K})^3$~cm.

\subsection{Neutrino yields and fluxes from a single source}

In order to analyze the effect of the thickness of the source on the
neutrino spectrum we study the neutrino yield of a single source. 
The neutrino yield $Y_\nu(E)$, i.e. the ratio $Y_\nu(E)=\phi_\nu(E)/(\tau
\phi_p(E))$ 
of the emitted neutrino flux $\phi_\nu$ and the product of the depth
$\tau$ and the injected proton flux $\phi_p$, represents the number of
neutrinos produced per injected proton with the  same energy. 
For the energy spectrum of the injected protons we assume a power 
law $dN/dE\propto E^{-\alpha}$ with $\alpha=2.2$, consistent with
typical predictions for Fermi shock acceleration. For illustration we
 choose the maximal energy of the initial proton 
spectrum as
$E_{\rm max}=10^{24}$~eV.

\begin{figure}[ht]
\begin{center}
\includegraphics*[width=0.45\textwidth,angle=0,clip]{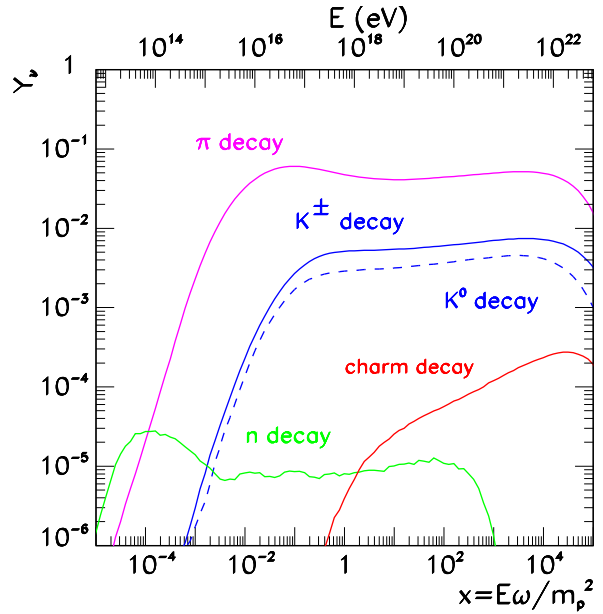}
\caption{\label {Y0} 
(Color online) Neutrino yield $Y_\nu$ as function of $x=E\omega/m_p^2$ from
the decay of $\pi^\pm$ in magenta (light gray), $K^\pm$ in solid blue
(dark gray), $K^0$ in dashed blue (dark gray),
neutrons in green (very light gray), and  charm mesons in red (gray)    
 for a source with $T=10^4$~K and $\tau=0.4$.}
\end{center}
\end{figure}

Let us first describe the different reactions contributing to the
neutrino flux. In Fig.~\ref{Y0}, we show the various contributions
to $Y_\nu$  for the case of an intermediate value of the 
interaction depth, $\tau=0.4$~\footnote{Throughout the present work we
will not distinguish neutrinos from antineutrinos. Therefore
$\phi_{\nu_\alpha}$ will represent the sum of $\phi_{\nu_\alpha} +
\phi_{\bar\nu_\alpha}$}.   
The low energy tail of neutrinos arises
from the decay of neutrons, $n\rightarrow p+e^-+\bar\nu_e$, where 
the energy fraction transferred to neutrinos is on average only 
$\sim 10^{-3}$. At all other energies,  charged pions are produced
most efficiently and their decay products provide the dominating contribution
to the neutrino yield. The various decay channels of kaons contribute 
around 10\% to the neutrino yields. The neutrino yield from prompt
decays of charm is even smaller, because of the relative high mass of
charm mesons resulting in a high energy threshold and small cross
section for charm production.

\begin{figure}[ht]
\begin{center}
\includegraphics*[width=0.45\textwidth,clip]{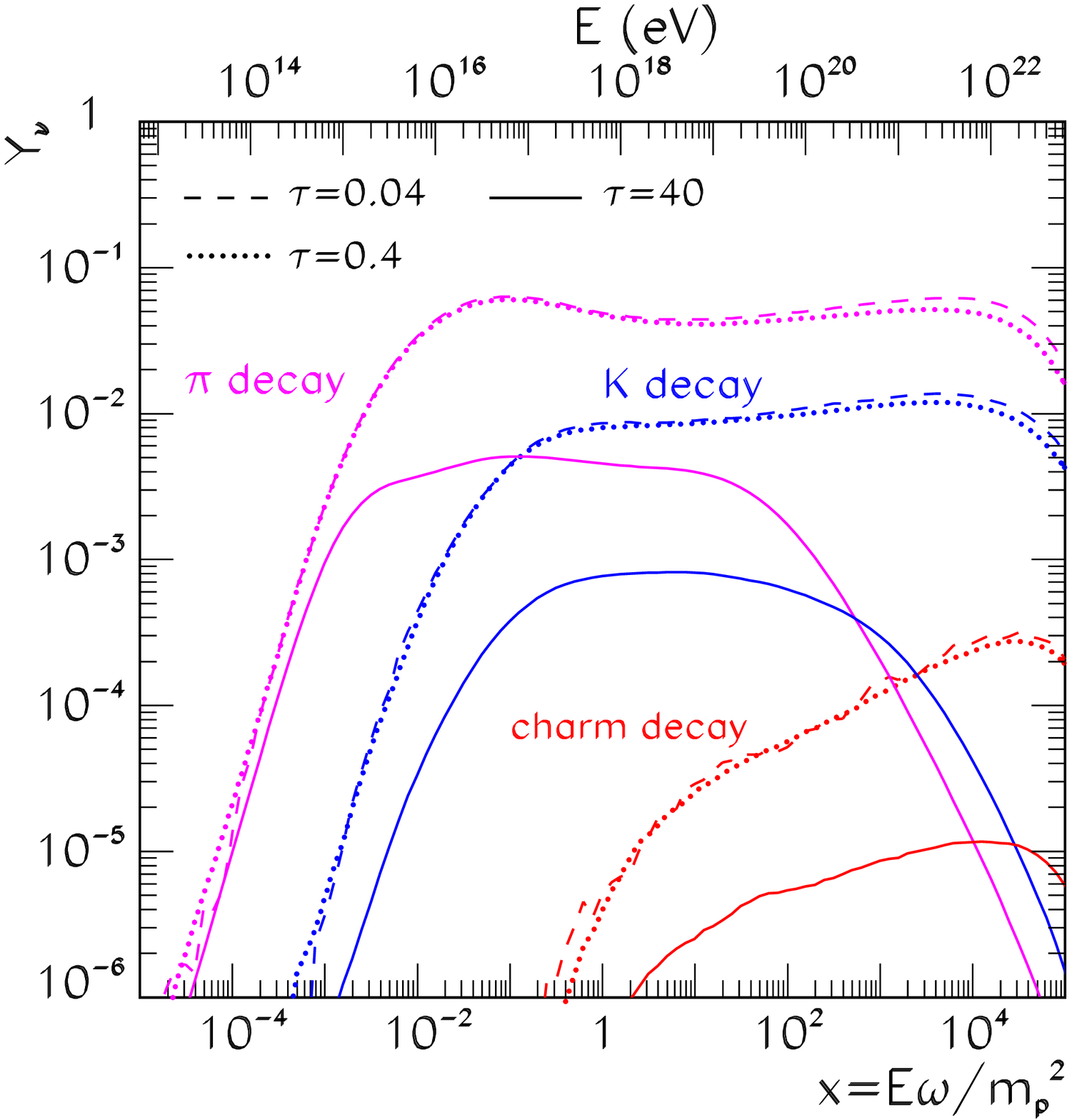}
\includegraphics*[width=0.45\textwidth,clip]{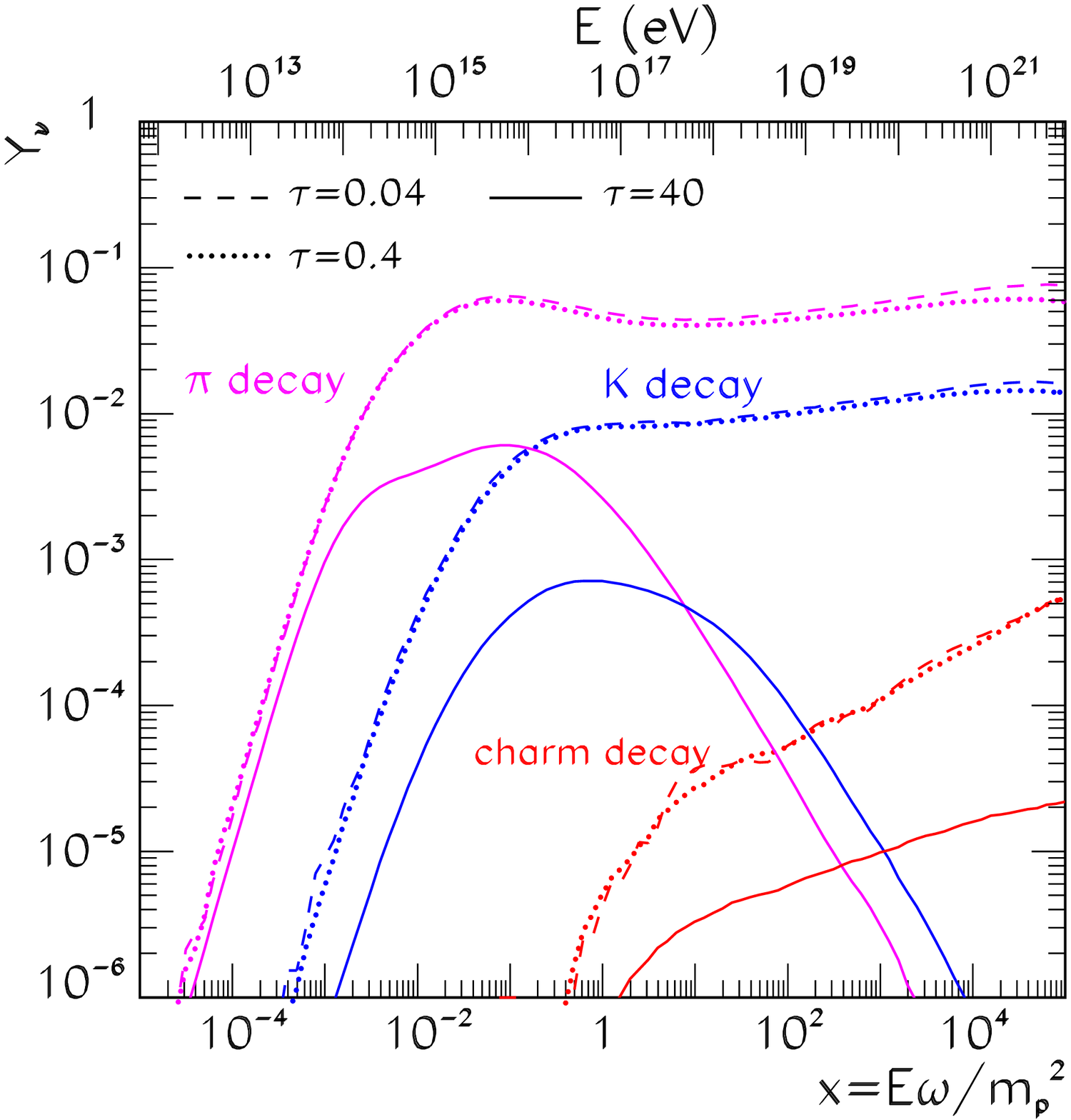}
\caption{\label{Y1} 
(Color online) Neutrino yield $Y_\nu$ as function of $x=E\omega/m_p^2$
  (bottom axis) and 
$E$ (upper axis) from pion magenta (light gray), 
kaon in blue (dark gray), and charm decays in red (gray) for
$\tau=0.04, 0.4$ and 40;  
upper panel for $T=10^4$~K, lower panel for $T=10^5$~K.}
\end{center}
\end{figure}

Next we discuss the effect of multiple scatterings on the different
neutrino yields. In the top panel of Fig.~\ref{Y1}, we show the
yields from meson decays for a source with $T=10^4$~K and three different 
sizes, $L=10^{13}$~cm ($\tau=0.04$),~$10^{14}$~cm ($\tau=0.4$) and
$10^{16}$~cm ($\tau=40$), respectively. For $\tau\lsim 1$,
multiple scattering is negligible, $\phi_\nu\propto\tau$ and the neutrino
yields expressed as functions of $x$ are thus independent from $\tau$. 
Moreover, the neutrino yield from charm meson decays is clearly subdominant. 
Increasing the size of the source,  multiple scattering cannot be
neglected anymore and three different $x$ regions can be
distinguished. For $x<x^\pi_{\rm cr}$, most pions decay before
scattering and, as in the case of transparent sources, the neutrino yield  
is dominated by the contribution of charged pions. In the intermediate
$x$ range, $x^\pi_{\rm cr}\lsim x \lsim x^K_{\rm cr}$, multiple
pion scattering on photons becomes effective and therefore
neutrinos from kaon decays start to provide the most important
contribution to the total neutrino yield~\footnote{It has been already
  pointed out that neutrinos from kaon decay can provide an important
  contribution to the neutrino signal in other astrophysical scenarios like
  GRBs~\cite{Ando:2005xi,Asano:2006zz}.}. 
Going to even higher
energies, $x\gsim x^K_{\rm cr}$, leads to a strong suppression of
neutrino from these decays as both pions and kaons most often scatter
before decaying. Hence at high energies,  neutrinos from
decays of charm mesons represent the main component of the total
neutrino yield. Moreover, the maximum of the neutrino yield does not
coincide with that of the transparent cases, because $\phi_\nu$ is not longer
proportional to the interaction depth.

In the bottom panel of Fig.~\ref{Y1}, we show again neutrino yields but
for a source with $T=10^5$~K. If multiple scattering
can be neglected, the neutrino yields are as expected independent from 
$T$ and $\tau$. For $\tau\gsim 1$,  we observe the same behavior of
the neutrino yields from pion and kaon decays as in the upper panel,
but the suppression of their fluxes starts already at lower energies. 
As anticipated in the discussion of Fig.~1, the range of energies
where neutrino fluxes are unsuppressed strongly depends on the energy
threshold $x_{\rm th}$ and the critical energy $x_{\rm cr}$. So,
whereas for a source at $T=10^4$~K the plateau visible in
Fig.~\ref{Y1} for $Y_\nu$  extends approximately over four orders of
magnitude, in the case of a source at $T=10^5$~K it is reduced by two
orders of magnitude.

\begin{figure}[ht]
\begin{center}
\includegraphics*[width=0.45\textwidth,angle=0,clip]{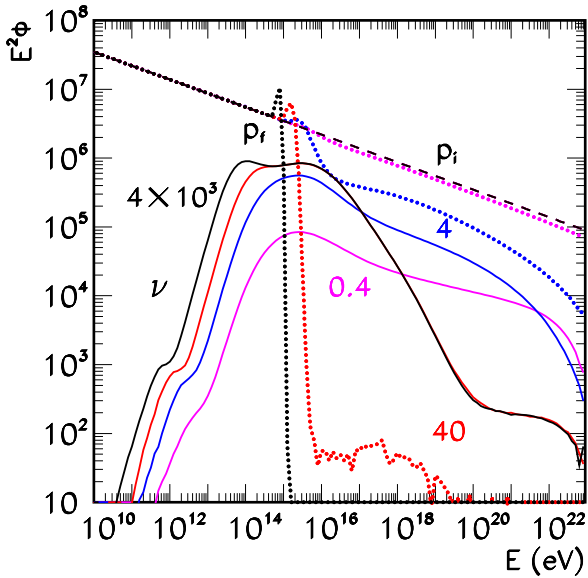}
\caption{\label{fluxes_tau} 
(Color online) Unnormalized fluxes of initial (dashed) and final (dotted) protons, as
well as the total neutrino (solid), for a source at $T=10^5$~K and with
interaction depths $\tau=0.4$ in magenta (light gray), 4 in blue (dark
gray), 40 in red (gray) and
$4\times 10^3$ in black.}
\end{center}
\end{figure}

\begin{figure}[ht]
\begin{center}
\includegraphics*[width=0.45\textwidth,angle=0,clip]{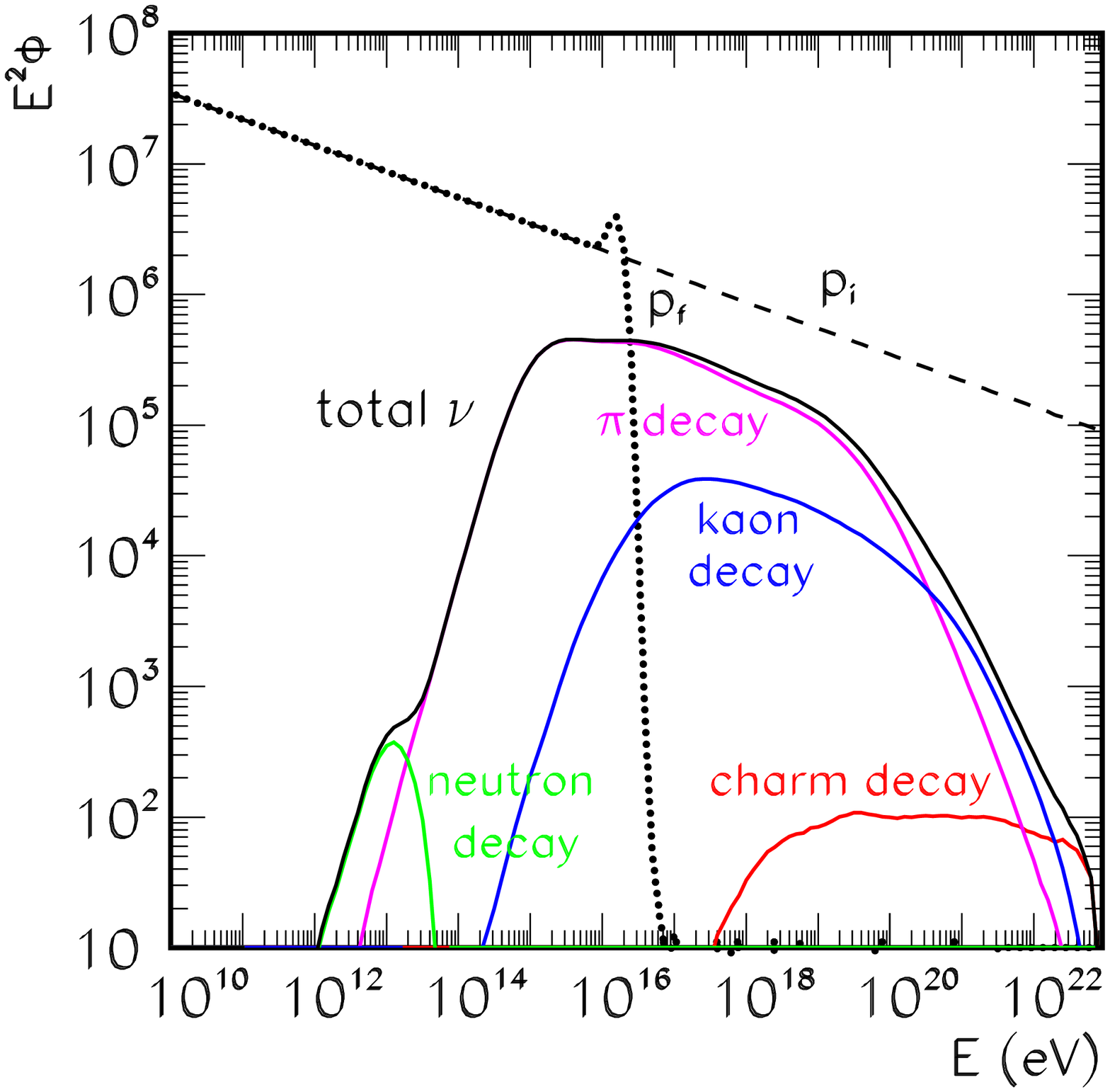}
\includegraphics*[width=0.45\textwidth,angle=0,clip]{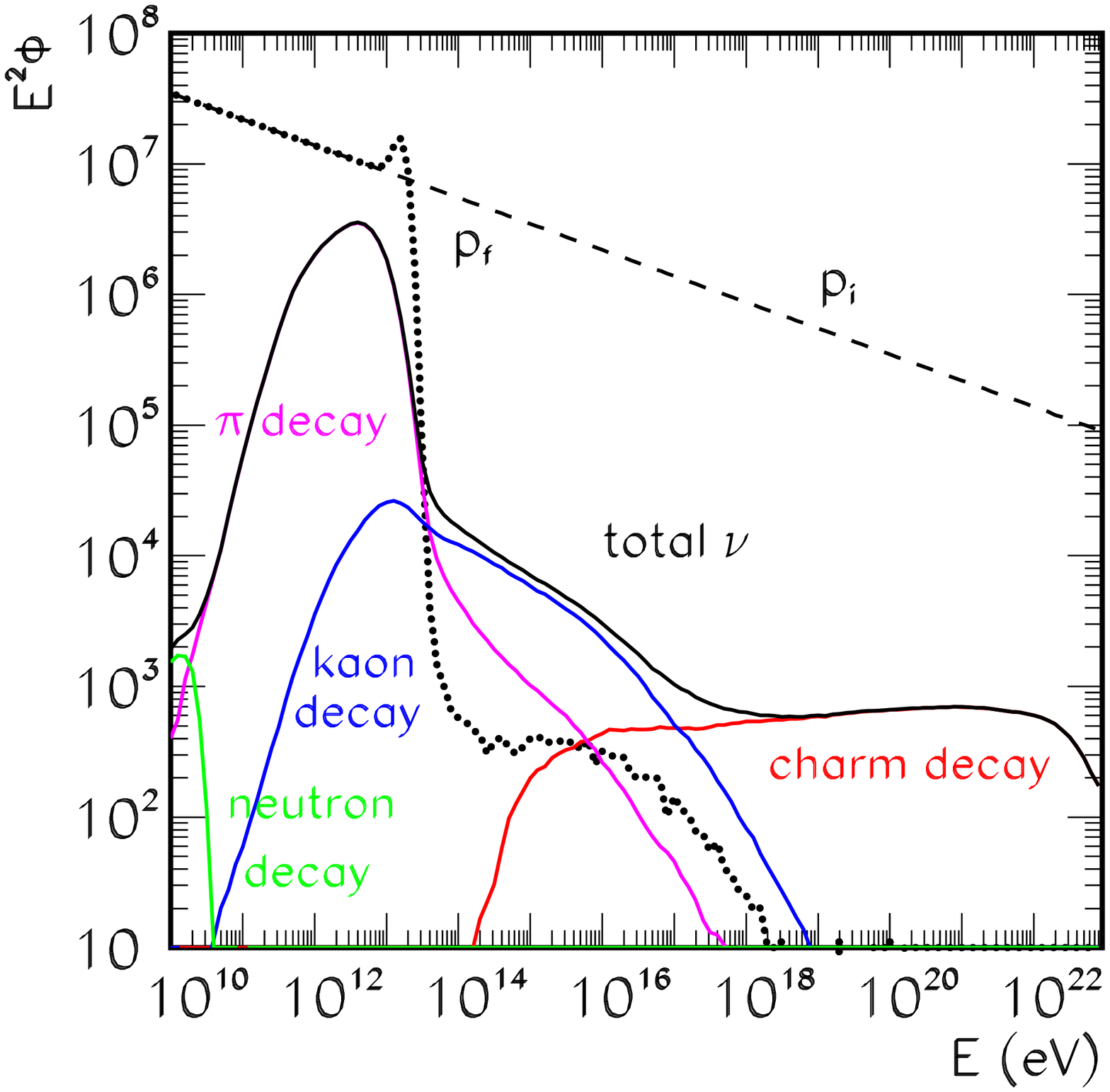}
\caption{\label{fluxes} 
(Color online) Unnormalized fluxes of initial (dashed) and final
  (dotted) protons, as 
well as the total neutrino (solid), for a source with interaction
depth $\tau=40$ at $T=10^4$~K (top) and $T=10^7$~K (bottom). The
different contributions to the total neutrino flux are also shown:
pion in magenta (light gray), kaon in blue (dark gray), charm meson in
red (gray), and neutron decay in green (very light gray). }
\end{center}
\end{figure}

Finally, we want to illustrate how the high-energy suppression of
neutrino fluxes depends on the interaction depth $\tau$, as well as on
 the density and the typical energy of
photons.  
Figure~\ref{fluxes_tau} shows (unnormalized) fluxes for a source at
$T=10^5$~K and different interaction depths, from 0.4 to $4\times
10^3$. In the case of thin sources, $\tau\lsim$ a few, we observe that
the final proton flux is only slightly distorted relative to the
initial flux, and therefore provides a non-negligible contribution to
the observed UHECRs. The neutrino flux in these sources is basically 
proportional to the depth. For opaque sources the final proton flux
becomes strongly suppressed  above threshold,
$E_p^{\rm th}\approx m_p m_\pi/(2\epsilon_\gamma)$. For such high
depths most pions and kaons scatter before decaying and as a
consequence the flux of neutrinos becomes also suppressed at energies
higher than $E_{\rm cr}$. The only effect of further
increasing $\tau$ is an increase of the low energy neutrinos, as the
high number of scatterings leads to more low energy mesons.
Figure~\ref{fluxes} shows (unnormalized) fluxes for an
opaque source, $\tau=40$, at $T=10^4$ K in the top and  $T=10^7$~K in
the bottom. 
As previously mentioned, the final proton flux is strongly suppressed
above threshold, whereas the  neutrino flux
rises already at lower energies, because they carry away only a
fraction of the proton energy. 
The energy range where the neutrino flux
is maximal extends approximately from  $E_p^{\rm th}$  
to $E_{\rm cr}$, which for the source at $T=10^4$~K, goes  roughly
from $10^{16}$~eV to 
$10^{20}$~eV. This range is  
strongly reduced for increasing $T$ as $E_{\rm cr}\propto 1/n\propto
1/T^3$, whereas $E_{\rm th}\propto 1/T$. 
Therefore, critical and threshold energies merge and only a a narrow
energy range around $E\approx 10^{12}$~eV with unsuppressed neutrino
flux remains for $T=10^7$~K, cf. the lower panel of Fig.~\ref{fluxes}.

\section{Flavor dependence of neutrino yields}

Both neutrino telescopes and extensive air shower experiments have
some flavor discrimination possibilities. The case of neutrino
telescopes is discussed for the example of ICECUBE in detail in
Ref.~\cite{Beacom:2003nh}. The long range of muons ensures that a muon
track from $\nu_\mu$ charged-current interactions is always visible
allowing the identification of these events. By contrast, the
charged-current interactions of $\nu_e$ and $\nu_\tau$ are---as long
as the tau decay length is too short to be detectable---only
distinguishable by the different muon content in electromagnetic and
hadronic showers. Therefore, these two flavors are in practice
impossible to differentiate 
for $E\lsim 5\times 10^{14}$~eV at a neutrino telescope. 
In the energy range 
$5\times 10^{14}{\rm eV}\lsim E \lsim 2\times 10^{16}{\rm eV}$,
the ``double-bang'' signature~\cite{double} 
of $\nu_\tau$ gives some handle for  the
identification of this flavor, while for higher energies at least one
of the two shower is outside a 1~km$^3$ detector. On the other hand,
extensive air shower experiments have the potential to identify
double-bang events in the small energy interval $5\times 10^{17}{\rm
  eV}\lsim E \lsim  2\times 10^{18}{\rm eV}$~\cite{EAS}. 
In summary, the main observable for the neutrino flavor composition is
the ratio of track to shower events in a neutrino telescope, 
$R_\mu=\phi_\mu/(\phi_e+\phi_\tau)$, while only in a very small energy
range all flavors can be distinguished. Additionally, extensive air
shower experiments are sensitive to the fraction of tau events in all
horizontal neutrino events, $R_\tau=\phi_\tau/(\phi_e+\phi_\mu)$, in
a small energy window around $10^{18}$~eV.

In spite of these flavor discrimination possibilities, the potential
of high energy neutrino observations for mixing parameter studies
are only rarely discussed (for some recent examples see Ref.~\cite{exc,exc2}).
One reason for this missing interest
is the small number of events expected in next generation
experiments like ICECUBE or AUGER even in optimistic scenarios.
Moreover, the neutrino flavor ratio from pion decay is 
$\phi(\nu_e):\phi(\nu_\mu):\phi(\nu_\tau)=1:2:0$ before
oscillations, resulting into
$\phi(\nu_e):\phi(\nu_\mu):\phi(\nu_\tau)=1:1:1$ at the detector quite
independent from unknown details of the neutrino mixing
matrix~\cite{double}. Thus 
the maximal mu-tau mixing together with the initial flavor ratio
$\phi(\nu_e):\phi(\nu_\mu):\phi(\nu_\tau)=1:2:0$ seems to disfavor
high energy neutrino experiments as tools to probe only poorly known
parameter as $\theta_{13}$ or completely unknown ones as the octant of
$\theta_{23}$ or the CP-violating phase $\dcp$ of the neutrino mixing
matrix.  As discussed in Refs.~\cite{exc,exc2}, there exist however
several examples of 
neutrino sources where at least in some energy range significant
deviations from this canonical flavor ratio can be expected. Such a
deviation does not only contain information about neutrino properties
but also their sources. For instance, Ref.~\cite{An04} discussed the 
possibility to distinguish between $pp$ or $p\gamma$ sources of
neutrinos measuring their flavor ratios. In the following, we will
show that opaque sources are characterized by an energy-dependent
flavor ratio and thus the flavor ration encodes non-trivial information.

\subsection{Energy dependence of the neutrino flavor ratio at the source}

\begin{figure}[ht]
\begin{center}
\includegraphics*[width=0.45\textwidth,angle=0,clip]{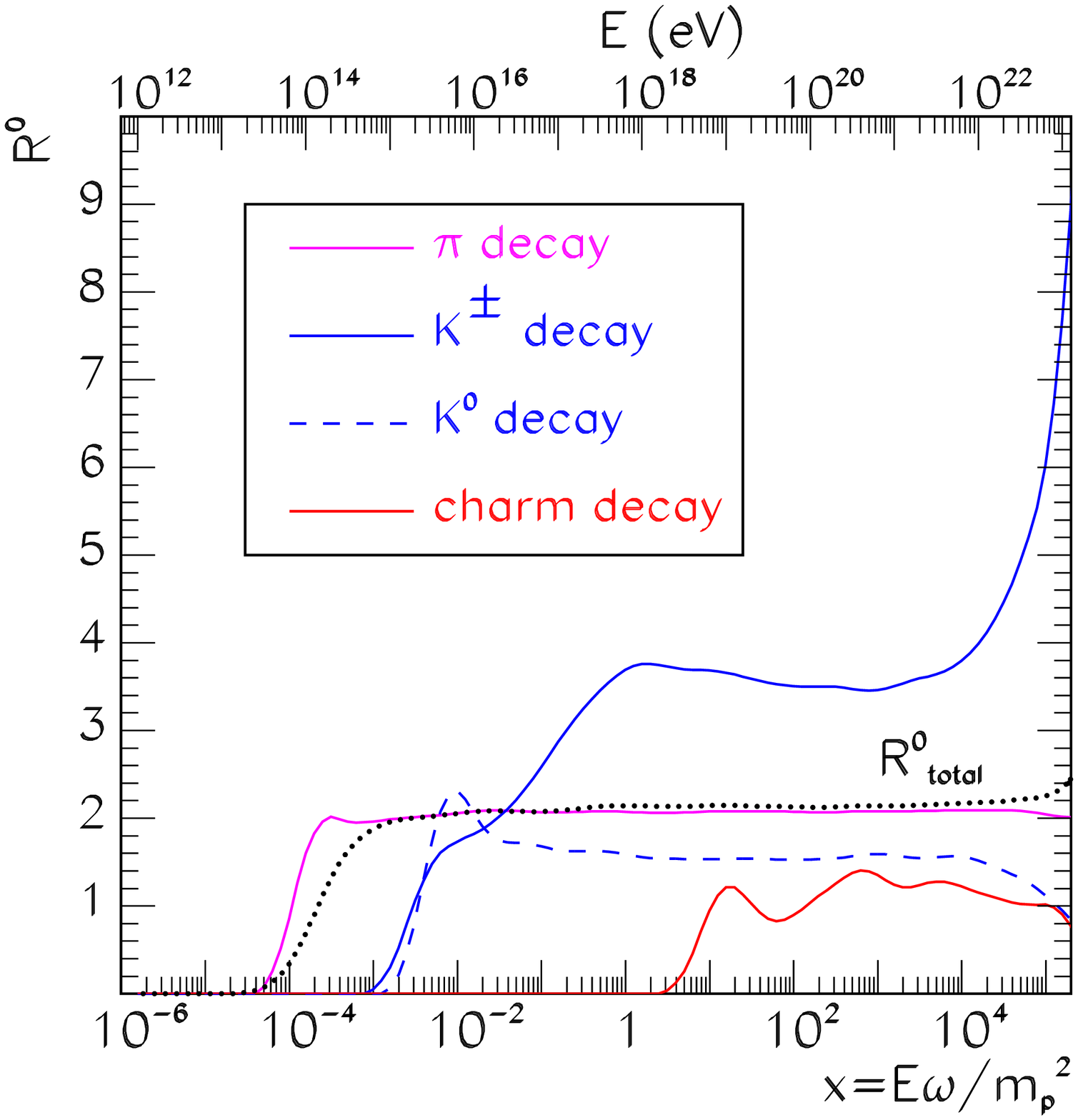}
\includegraphics*[width=0.45\textwidth,angle=0,clip]{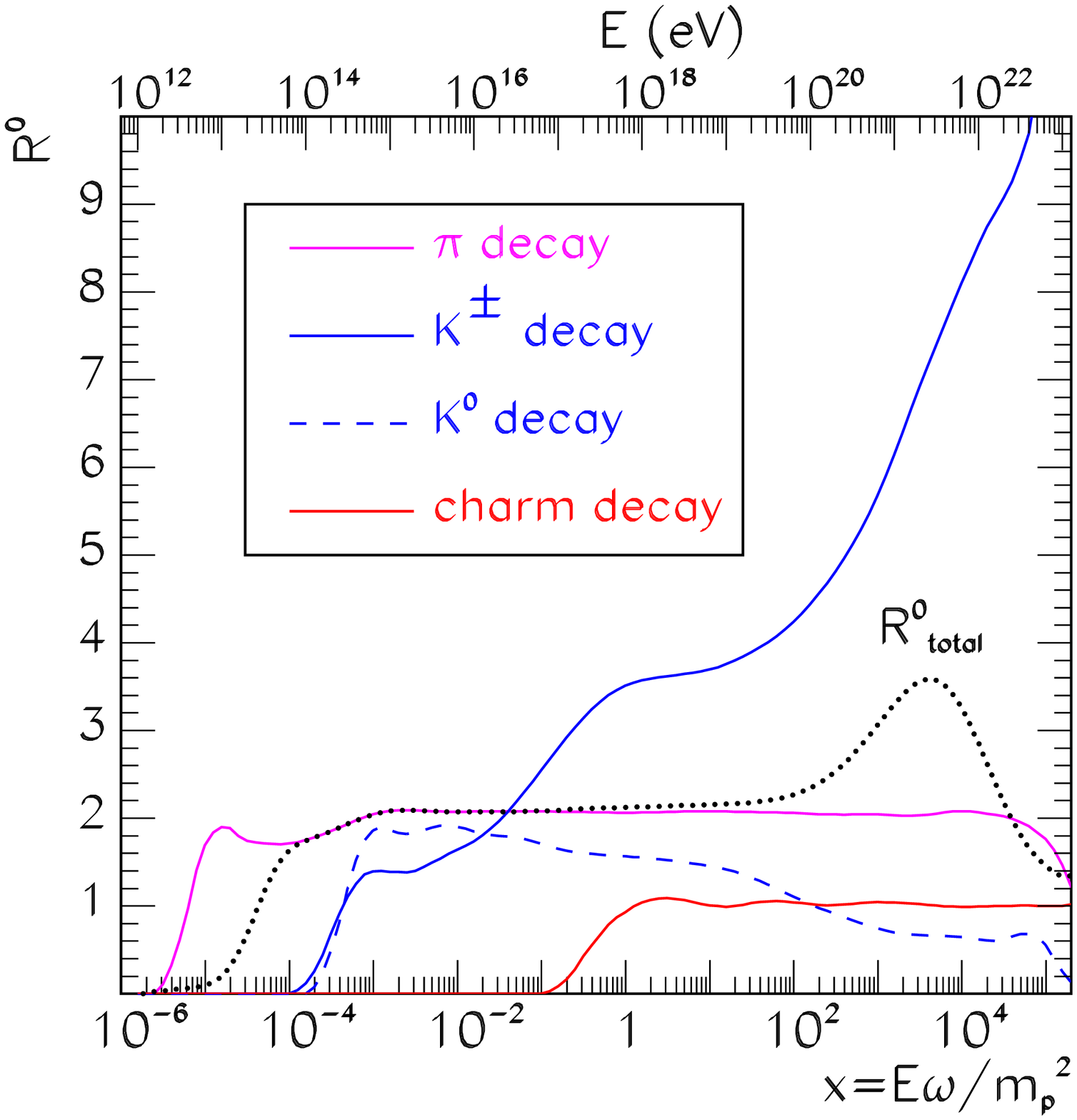}
\caption{\label {Yflavour1} 
(Color online) Flavor ratio $R^0$ of the neutrinos produced in a source before they
oscillate in terms of $x=E\omega/m_p^2$. Top: Specific flavor ratio
for the neutrinos generated 
from the decay of $\pi^\pm$ in magenta (light gray), $K^\pm$ in solid
blue (dark gray),
$K^0$ in dashed blue (dark gray),  and  charm meson in red (gray)     
 for a source with $T=10^4$~K and $\tau=0.04$. In black dotted line is
 shown the total $R^0$.
Bottom: The same for a source with interaction depth $\tau=40$.}
\end{center}
\end{figure}

We start with an analysis of the expected neutrino flavor ratio
$R^0\equiv Y_{\nu_\mu}/Y_{\nu_e}$ at the source. In
Fig.~\ref{Yflavour1}, we show $R^0$ separately for 
each reaction contributing to the neutrino yield in the case of a
source with $T=10^4$~K. For definiteness we have set $R^0=0$, if
$\phi_\nu(x)/\phi_p(E)<10^{-6}$. The two panels show $R^0$ for
transparent source with interaction depth $\tau=0.04$ (top) and
$\tau=40$ (bottom), respectively .   

In pion decays, all three neutrinos produced have very similar energy
spectra and therefore $R^0$ is close to two. The most important
decay channel of charged kaons is the same as the one of charged pions, 
$K^\pm \to \mu + \nu_\mu \to 2\nu_\mu + \nu_e + e$. However, the mass
difference between kaons and muons leads to very different energy
distributions of the directly produced $\nu_\mu$ neutrinos. The
neutrino spectra from kaons are dominated at low energies by neutrinos
from muon decay, whereas at high energies $\nu_\mu$'s produced in the
direct kaon decays dominate. As a consequence, there are three
different ranges for the flavor ratio $R^0$ of neutrinos 
produced by the charged kaon decays: At low energies the contribution of
the direct $\nu_\mu$ is negligible.  Thus $R^0$ approaches one at low
energies, since the $\nu_\mu$ and $\nu_e$ spectra from muon decay are similar.
Near the threshold energy for kaon production, $x^{K}_{\rm th}\approx
0.2$, direct $\nu_\mu$'s start to dominate, leading to an increase of $R^0$ to
values larger than two. The particular value of $R^0$ within this
energy range depends on the initial
kaon spectrum: the steeper it is, the larger the flavor ratio. 
As the energy increases the neutrino flavor ratio remains roughly
constant until it gets closer to the high energy kinematical limit. At
this point there are practically only $\nu_\mu$'s from the direct kaon
decay and therefore $R^0$ increases fast. 
In the case of neutrinos produced in neutral kaon decays one expects a
flavor ratio between two, corresponding to the contribution from
$K^0_S$ (via $K^0_S\rightarrow 2\pi$), and roughly one, due to the
neutrino yield from the $K^0_L$ decay, mainly from $K^0_L\rightarrow
\pi+e+\nu_e$. 
Finally, concerning the neutrinos generated in the decay of charm
mesons  one observes the same contribution from $\nu_e$ and
$\nu_\mu$, leading to a flavor ratio of one~\footnote{The $\nu_\tau$
  production is negligible in comparison to the other flavors.}. 

The flavor ratio for an opaque source with $\tau=40$ (bottom panel) 
is only changed for neutrinos from kaon decays.
At energies larger than $x_{\rm cr}^K$ most of charged kaons scatter
before decaying, leading to a steeper spectrum of the kaons which
eventually decay. As a consequence at that point
the value of $R^0$ increases a second time before approaching the high
energy limit.
Roughly at the same energies the pions emitted by $K^0_L$  scatter
before decaying. Therefore, in contrast to transparent sources,
the flavor content of  $K^0_L$'s at high energies is  dominated by
$\nu_e$ and the ratio $R^0$ tends to values smaller than one.

\begin{figure}[ht]
\begin{center}
\includegraphics*[width=0.45\textwidth,angle=0,clip]{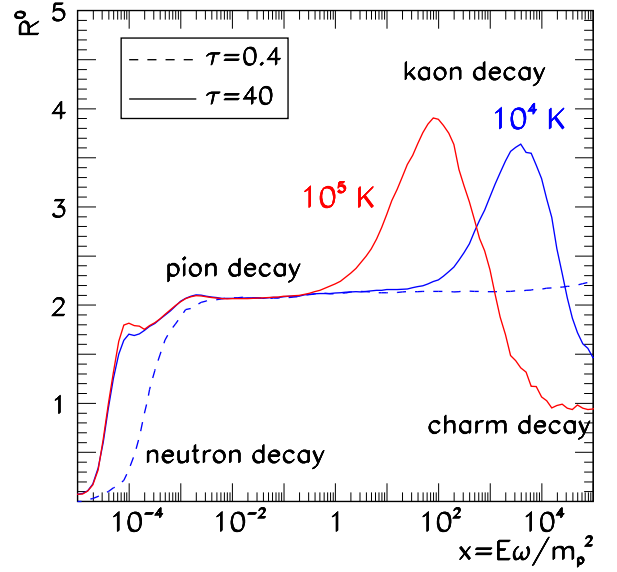}
\caption{\label {Yflavour2} 
(Color online) Total flavor ratio for a source at $T=10^4$~K  and $\tau=0.4$
in dashed blue (dark gray), $T=10^4$~K and 
$\tau=40$ in solid blue (dark gray), and $T=10^5$~K and
$\tau=40$ in solid red (gray). The main reactions giving rise to the neutrino
signal are also shown.}
\end{center}
\end{figure}

Let us now consider the total flavor ratio expected for different
sources, shown in the  Fig.~\ref{Yflavour2}. 
In the case of a transparent source the value of $R^0$ is small at
low $x$ because almost all neutrinos are produced in the decay of
neutrons, see Fig.~\ref{Y0}. Once the contribution from
charged pion decay becomes dominating, $R^0$ approaches the standard
value two, as expected from the upper panel of Fig.~\ref{Yflavour1}.  
In contrast to transparent sources, 
in the case of an opaque source, $\tau=40$, $R^0$ is strongly
energy-dependent. In particular, the flavor ratio
reaches the value of
2 at lower $x$, because multiple scattering of nucleons leads to lower
energies of the escaping neutrons. Moreover, there is a bump in the
range $100\lsim x \lsim 10^4$ and $1\lsim x\lsim 100$ for
the case $T=10^4$~K and $10^5$~K, respectively. The lower limit of
theses ranges corresponds to the cross-over between pion and kaon
dominance, while the upper limit corresponds to the cross-over between
kaon and charm dominance. As can be inferred from the bottom panel of
Fig.~\ref{Yflavour1}, the value of $R^0$ at the bump is determined by
the interplay of the charged and neutral kaon neutrinos.
Finally, in both cases the asymptotic value is $R^0\sim 1$,
which corresponds to the ratio expected when the decay of charm mesons
dominate.

\subsection{Effect of neutrino oscillations}

The neutrino spectra at the source discussed in the preceding
subsection are modulated by oscillations. Therefore the expected
flavor ratios $R_i$ at the Earth are different from the original
ratios $R_i^0$ at the source. 

\begin{figure}[ht]
\begin{center}
\includegraphics*[width=0.45\textwidth,angle=0,clip]{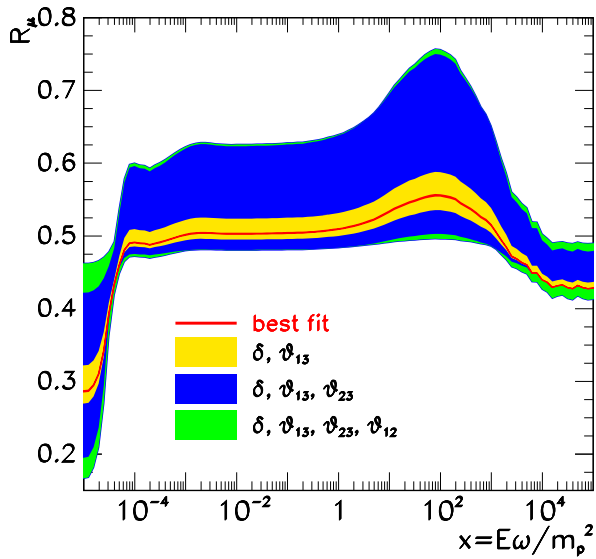}
\caption{\label {Yflavourosc} 
(Color online) Flavor ratio $R_\mu$ at the Earth for a source with $T=10^5$~K and
$\tau=40$. The solid red line (gray) corresponds to the best-fit point of the
neutrino mixing parameters, $\sin^2\theta_{12}=0.3$, $\sin^2\theta_{23}=0.5$ 
and $\theta_{13}=0$. In the colored areas mixing angles within 
95\%~C.L. and a possible non-zero value of $\delta_{\rm CP}$ have been
considered: $\theta_{13}$ and $\delta_{CP}$ in yellow (very light gray), plus
$\theta_{23}$ in blue (dark gray), and  $\theta_{12}$ in green
(light gray).} 
\end{center}
\end{figure}

The neutrino fluxes arriving at the detector, $\phi^D_\alpha$, can be written
in terms of the initial fluxes $\phi_\alpha$ and the 
conversion probabilities $P_{\alpha\beta}$,
\begin{equation}
\phi^D_\alpha = \sum_\beta P_{\alpha\beta}\phi_\beta = P_{\alpha
  e}\phi_e + P_{\alpha \mu}\phi_\mu \,.
\end{equation}
Since the interference terms sensitive to the mass splittings $\Delta m^2$'s
do not contribute, the conversion probabilities are simply
\begin{equation}
P_{\alpha\beta} = \delta_{\alpha\beta}-2\sum_{j>k}\Re(U^\star_{\beta
  j}U_{\beta k}U_{\alpha j}U^\star_{\alpha k}) \,,
\end{equation}
where $U$ is the neutrino mixing matrix and Greek (Latin) letters are
used as flavor (mass) indices.
In Fig.~\ref{Yflavourosc}, we show the expected flavor ratio $R_\mu$ at the
detector in the case of a source with $T=10^5$~K and $\tau=40$ as a
solid, red line for the best-fit point of the neutrino mixing
parameters, $\sin^2\theta_{12}=0.3,~\sin^2\theta_{23}=0.5$ and
$\theta_{13}=0$~\cite{nufit}. Various bands show the range of $R_\mu$
allowed if the mixing parameters are varied within their  95\%~C.L.,
while the unconstrained CP phase $\delta_{\rm CP}$
is  varied in the whole possible range, $\delta_{\rm CP}\in[0:\pi]$.
Dominant uncertainty for the prediction of $R_\mu$ is the
value of $\theta_{23}$. 

We analyze next the dependence of the flavor ratio $R_\mu$ on the
different neutrino mixing parameters in more detail.  Expanding
$R_\mu$ up to first order in $\theta_{13}$ and choosing for
illustration $\theta_{12}=\pi/6$ gives
\begin{eqnarray}
\lefteqn{\!\!\!\!\!
 R_\mu(R^0,\theta_{23},\theta_{13},\delta_{CP}) = \frac{P_{e\mu}+R_0
  P_{\mu\mu}}{P_{ee}+R_0 P_{e\mu}+P_{e\tau}+R_0 
  P_{\mu\tau}} } \nonumber\\ 
&\! = & \!\!\! A(R^0,\theta_{23}) +
  B(R^0,\theta_{23})\cos\delta_{\rm CP}\theta_{13} + O(\theta_{13}^2) .
\label{Rexpansion}
\end{eqnarray}
The explicit expression for $A$ and $B$ are given in the Appendix.
In Fig.~\ref{A-B}, we show the dependence of the coefficients $A$
(left panel) and $B$ (right panel) on $R^0$ and $\theta_{23}$.  
While the parameter $A$ is more sensitive to the value of $R^0$  if
$\theta_{23}$ is in the second octant, $\sin^2\theta_{23}>0.5$, 
this behavior is opposite for $B$.
\begin{figure}[ht!]
\begin{center}
\includegraphics*[width=0.45\textwidth,angle=0,clip]{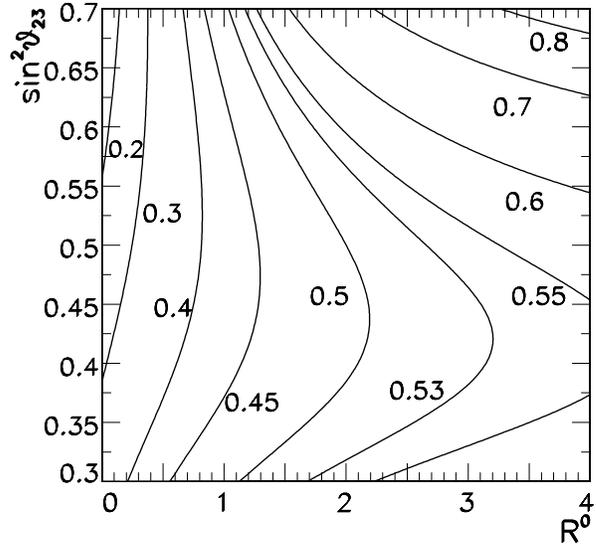}
\includegraphics*[width=0.45\textwidth,angle=0,clip]{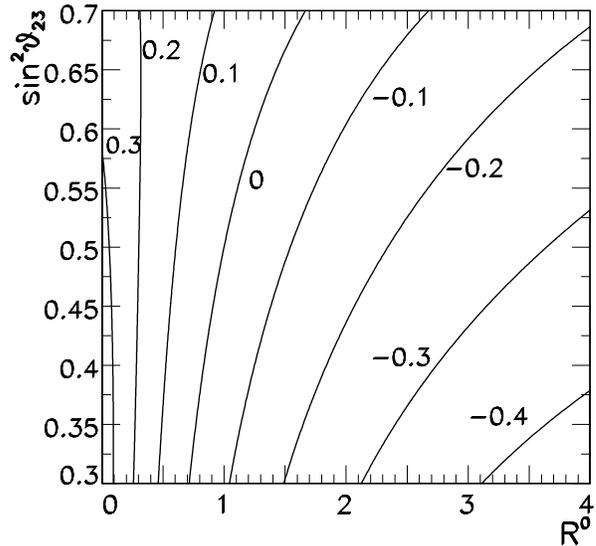}
\caption{\label {A-B} 
Contours of the coefficients $A(R^0,\theta_{23})$ (top) and $B(R^0,\theta_{23})$
(bottom) in terms of $R^0$ and $\sin^2\theta_{23}$.}
\end{center}
\end{figure}

\begin{figure}[ht!]
\begin{center}
\includegraphics*[width=0.45\textwidth,angle=0,clip]{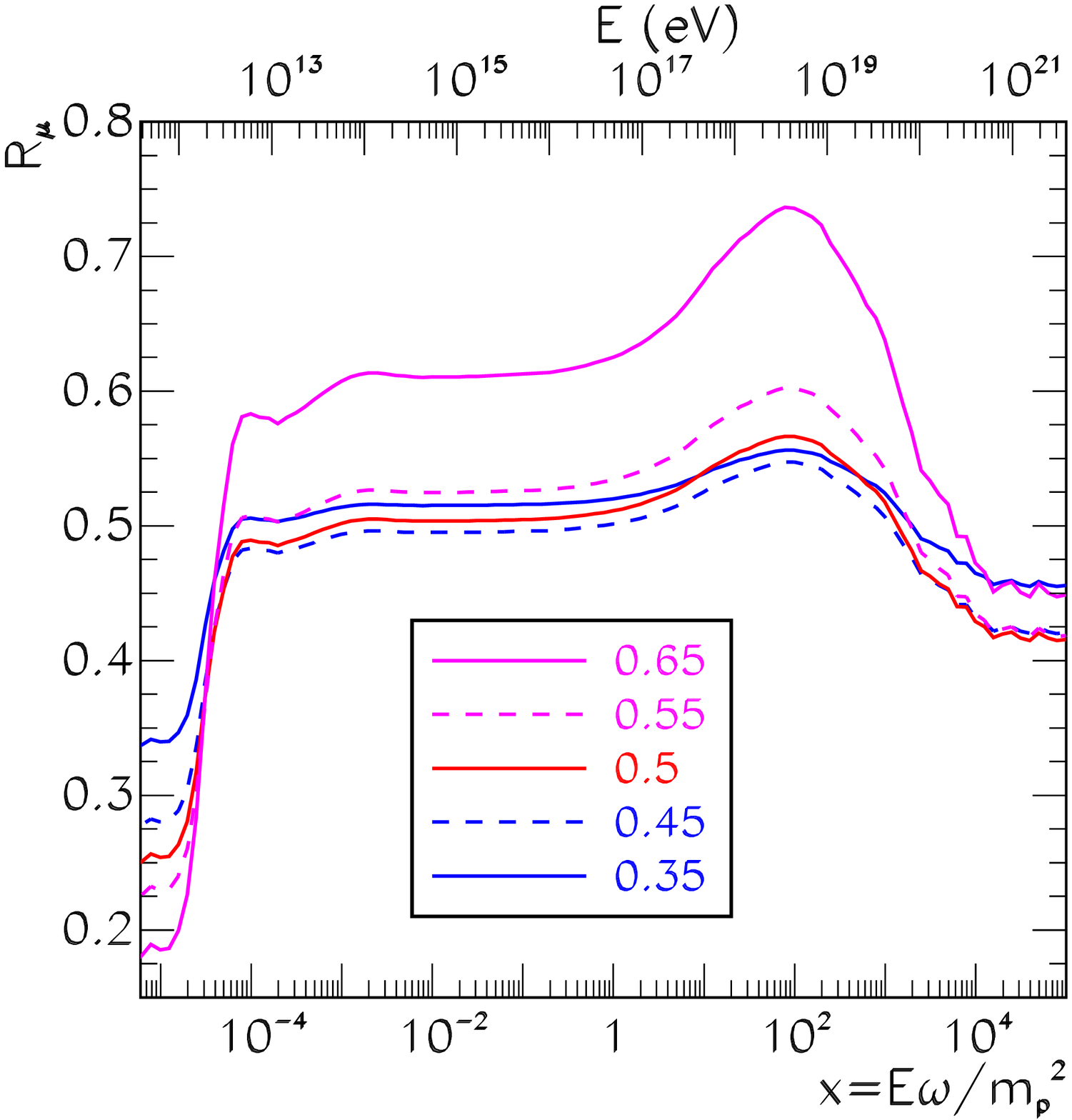}
\includegraphics*[width=0.45\textwidth,angle=0,clip]{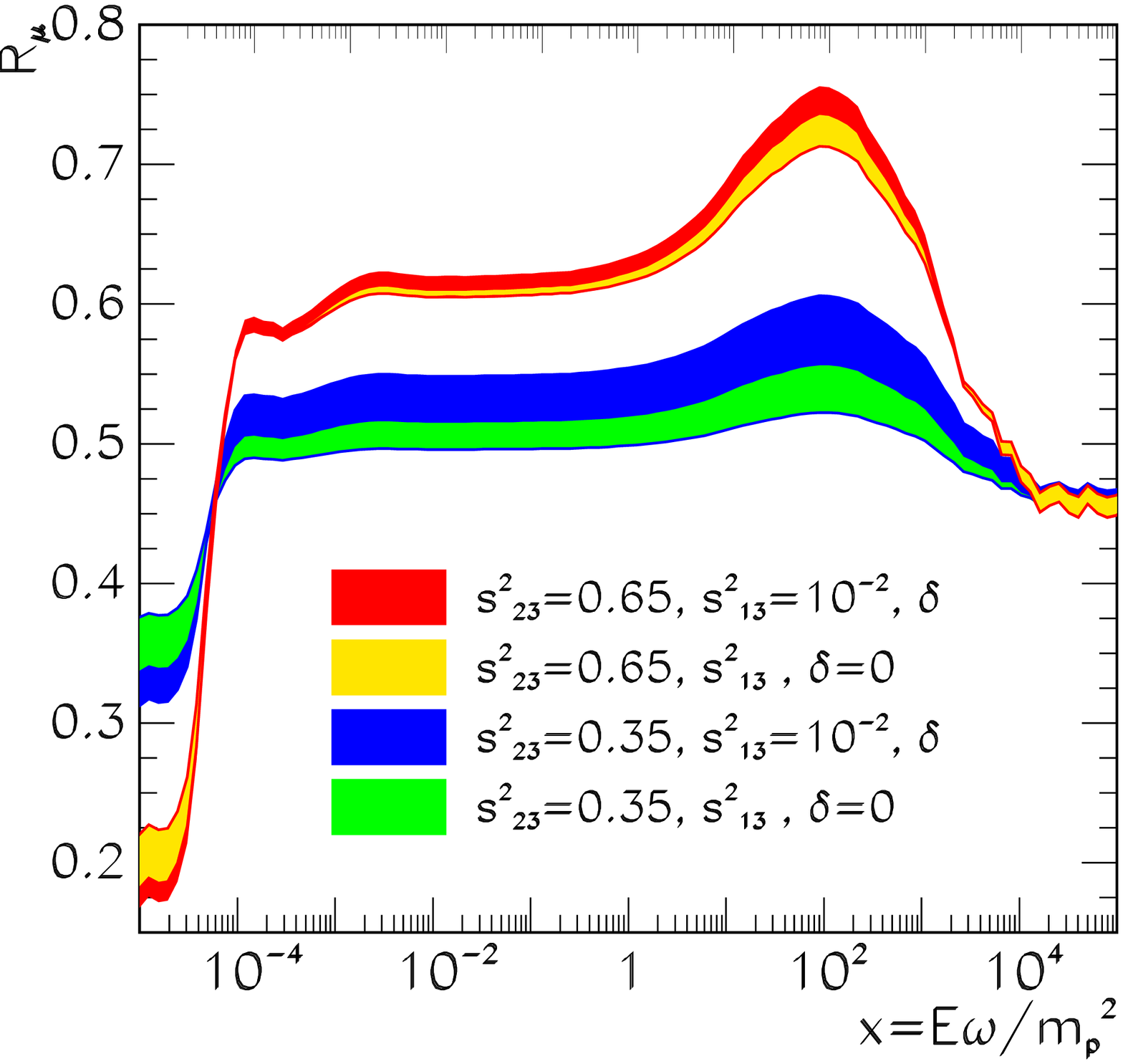}
\caption{\label {R_s223} 
(Color online) Top: Flavor ratio at the Earth $R_\mu$ for different values of
$\sin^2\theta_{23}$ assuming $\sin^2\theta_{12}=\pi/6$ and
$\theta_{13}=0$ for a source at $T=10^5$~K and interaction depth
$\tau=40$.
Bottom: Dependence of $R_\mu$ on $\theta_{13}$ for $\delta_{CP}=0$
in yellow (very light gray) and green (light green),and on $\delta_{CP}$, assuming
$\sin^2\theta_{13}=10^{-2}$ in red (gray) and blue (dark gray), for two extreme
 values of $\sin^2\theta_{23}$.
}
\end{center}
\end{figure}

Let us first  analyze the dependence of $R_\mu$ on $\theta_{23}$. For the
sake of simplicity, we assume $\theta_{13}=0$, and thus
$R_\mu(R^0,\theta_{23},0,\delta_{CP})=A(R^0,\theta_{23})$.
The upper panel of Fig.~\ref{R_s223} shows the flavor ratio $R_\mu$
for different 
values of $\sin^2\theta_{23}$. The possibility to extract information
about the value of $\theta_{23}$ strongly depends on the energy
range considered. At low energies, where the initial ratio is close to
zero (neutron neutrinos), $R_\mu$  depends inversely on
$\sin^2\theta_{23}$, varying between 0.15 and 0.35 for
$\sin^2\theta_{23}=0.65$ and $0.35$, respectively. This large
sensitivity of $R_\mu$ to the value of $\theta_{23}$  makes the
neutrinos emitted in neutron decay an alternative tool to study
this mixing angle~\cite{exc2}. 
At intermediate energies the sensitivity becomes smaller.  
When neutrinos come mainly from charged pion decays, $R^0\approx
2$, we observe basically two regions: below $\sin^2\theta_{23}\simeq
0.52$, the flavor ratio at the Earth lies around 0.5, whereas for
larger angles it grows until values around 0.6. This is a consequence
of the flat behavior of $A$ at $R^0$ larger than one for $\theta_{23}$
in the first octant, see the upper panel of
Fig.~\ref{A-B}. This difference can even increase at the bump, where
$R_\mu$ can reach values larger than 0.7 for $\theta_{23}$ in the second
octant whereas it remains smaller than 0.6 for angles in the first
octant. 
 Finally at higher energies (charm neutrinos) it becomes
very difficult to disentangle the different values: $R^0\simeq 1$ and
$R_\mu$ becomes roughly 0.45 for all $\theta_{23}$ around $\pi/4$, see upper
panel of Fig.~\ref{A-B}.

Let us now briefly discuss the dependence of $R_\mu$ on $\theta_{13}$
and $\delta_{\rm CP}$. According to Eq.~(\ref{Rexpansion}),
this variation depends on $R^0$ and $\theta_{23}$ through the function
$B(R^0,\theta_{23})$. The  bottom panel of Fig~\ref{A-B} shows that the
maximal variation takes place when $\theta_{23}$ is in the
first octant and $R^0$ deviates significantly from one.
In all cases, though, the sensitivity does not exceed 10\%.
In the bottom panel of Fig.~\ref{R_s223}, we show the variation of $R$
for values  of $\theta_{13}$ between 0 and $10^{-2}$, assuming 
$\delta_{\rm CP}=0$ and $\sin^2\theta_{23}=0.35$ (green) and 0.65
(yellow). While at low energies ($R^0\sim 0$) both cases give similar
results, at higher $R^0$ the dependence on $\theta_{13}$ is almost
negligible. 
We show also the variation of $R_\mu$ as function of $\delta_{\rm CP}$ for
the same values of $\sin^2\theta_{23}$ and
$\sin^2\theta_{13}=10^{-2}$. In this case, the variation is larger due
to the two different signs of $\cos\delta_{\rm CP}$. 

\section{Summary}

The high-energy cosmic ray flux from opaque or hidden neutrino sources
is suppressed  and these source are therefore among the most promising
candidates for powerful neutrino sources. 
We have calculated the yield of high energy neutrinos
from sources with photons as target material, paying especially attention 
to opaque sources.
We have found that the neutrino spectra 
from meson and muon decays are strongly modified with respect to transparent 
sources as soon as multiple scattering of nucleons becomes important.
The main consequence is a strong suppression of the neutrino flux
 from pion and kaon decays at energies above a critical energy
$E_{\rm cr}$, defined as the energy at
which the scattering length of mesons equals the decay length. Above
this energy both pions and kaons scatter before decay and thus
 the main contribution to the neutrino flux is supplied by
the decay of charm mesons.
As a result, the parameter range where opaque
sources can produce ultra-high energy neutrinos is rather restricted. 
An opaque source with $E_{\rm cr}\gsim 10^{21}$~eV requires basically
a large extension with low densities.

Both neutrino telescopes and extensive air shower experiments can not
only detect high energy neutrinos but also they have
some flavor discrimination possibilities. 
Since the determination of the neutrino flavor ratio can shed light on the 
properties of both sources and neutrinos, we have calculated  also
the expected flavor ratio for opaque sources containing photons as 
scattering targets.
The main
characteristic is a strong energy-dependence of the ratio $R^0$ of 
$\nu_\mu$ and $\nu_e$ fluxes at the sources. A generic  prediction for
neutrinos from hadron-photon scattering is the flavor ratio $R^0\sim
0$  below the threshold, i.e.\ a flux dominated by $\bar\nu_e$ at the
source. In the case of opaque sources the flavor ratio presents a
bump close to the energy where the decay length of charged 
pions and kaons equals their interaction length on target
photons.
The neutrino spectra at the source are modulated by
oscillation. Therefore the expected flavor ratio $R_\mu$ at the Earth
will be different from the original one $R^0$: it significantly
depends on the neutrino mixing parameters, and especially on $\theta_{23}$.
Therefore the observation of a strong energy dependence of $R_\mu$
will not only be a hint on the kind of source but also allow obtaining
complementary information on the neutrino mixing angles.

\section*{Acknowledgments}

We would like to thank Pedro Ru\'{\i}z Femen\'{\i}a and in particular 
Sergey Ostapchenko for useful and pleasant discussions. RT was
supported by the Juan de la Cierva programme, an ERG from the European
and by the Spanish grant FPA2005-01269. We would also like to thank the
European Network of Theoretical Astroparticle Physics ILIAS/N6 under
contract number RII3-CT-2004-506222.

\begin{widetext}
\appendix
\section*{Appendix: Two two-column formulae}
The explicit expression for $A$ and $B$ are 
\begin{eqnarray}
A(R^0,\theta_{23}) & = &
\frac{12+39R^0-12(R^0-1)
  \cos(2\theta_{23})+13R^0
  \cos(4\theta_{23})}{52+25R^0+12(R^0-1)\cos(2\theta_{23})-13R^0\cos(4\theta_{23})}     
 \\ 
B(R^0,\theta_{23}) & = &
\frac{-512\sqrt{3}(1+R^0)(-1+R^0+R^0\cos(2\theta_{23}))\sin(2\theta_{23})}{[
  52+25R^0+12(R^0-1)\cos(2\theta_{23})-13R^0\cos(4\theta_{23})]^2}. 
\end{eqnarray}
\end{widetext}

\end{document}